\documentclass[aps,pre,twocolumn,showpacs,floatfix,groupedaddress,nofootinbib]{revtex4}
\usepackage{algorithm}
\usepackage{algpseudocode}
\usepackage{amsmath}
\usepackage{amsthm}
\usepackage{amssymb}
\usepackage{amscd}
\usepackage{bm}
\usepackage{latexsym}
\usepackage{mathtools}
\usepackage{comment}

\usepackage{graphicx}
\usepackage{epsfig}
\usepackage{epstopdf}
\usepackage{svg}
\usepackage[caption=false]{subfig}
\usepackage{placeins}
\usepackage{tabularx}

\usepackage{url}
\usepackage{hyperref}

\renewcommand{\vec}[1]{\boldsymbol{#1}}
\providecommand{\mat}[1]{\mathbf{#1}}

\providecommand{\config}[1]{\mathbf{#1}}

\providecommand{\diff}{\mathrm{d}}

\providecommand{\norm}[1]{\Vert#1\Vert}

\providecommand{\config}{{\bf C}}
\providecommand{\eqref}[1]{(\ref{#1})}

\providecommand{\etal}{et al.\ }

\providecommand{\vanish}[1]{}

\newtheorem{theorem}{Theorem}
\newtheorem{conjecture}[theorem]{Conjecture}

\begin{document}
	
	\title{A geometric conjecture about phase transitions}
	
	\author{O. B. Eri\c{c}ok}
	\email{oericok@ucdavis.edu}
	\affiliation{Materials Science and Engineering, University of California, Davis, CA, 95616, USA.}
	
	\author{J. K. Mason}
	\email{jkmason@ucdavis.edu}
	\affiliation{Materials Science and Engineering, University of California, Davis, CA, 95616, USA.}
	
	\begin{abstract}
		As phenomena that necessarily emerge from the collective behavior of interacting particles, phase transitions continue to be difficult to predict using statistical thermodynamics.
		A recent proposal called the topological hypothesis suggests that the existence of a phase transition could perhaps be inferred from changes to the topology of the accessible part of the configuration space.
		This paper instead suggests that such a topological change is often associated with a dramatic change in the configuration space geometry, and that the geometric change is the actual driver of the phase transition.
		More precisely, a geometric change that brings about a discontinuity in the mixing time required for an initial probability distribution on the configuration space to reach steady-state is conjectured to be related to the onset of a phase transition in the thermodynamic limit.
		This conjecture is tested by evaluating the diffusion diameter and $\epsilon$-mixing time of the configuration spaces of hard disk and hard sphere systems of increasing size.
		Explicit geometries are constructed for the configuration spaces of these systems, and numerical evidence suggests that a discontinuity in the $\epsilon$-mixing time coincides with the solid-fluid phase transition in the thermodynamic limit.
	\end{abstract}
	
	
	\maketitle

	\section{Introduction}
	\label{sec:introduction}
	
	Phase transitions are essential to a variety of scientific and engineering applications, but continue to be difficult to predict from fundamental considerations.
	Instead, a phase transition is usually identified by an observed discontinuity in one of the derivatives of a thermodynamic potential.
	There have been several proposals concerning the origin of these discontinuities.
	For example, Landau theory \cite{landau1936theory,landau2013statistical} associates first-order phase transitions with spontaneous symmetry breaking as quantified by an appropriately-constructed order parameter.
	Such order parameters often need to be defined ex post facto though, after the characteristics of the phases involved in the phase transition are already known.
	An understanding of phase transitions that derives from more fundamental considerations would therefore be valuable.
	
	Statistical thermodynamics suggests that thermodynamic observables can be calculated as time averages of the relevant microscopic quantity along the system's trajectory through the phase space.
	The ergodic hypothesis \cite{boltzmann1871einige,ehrenfest1990conceptual} implies that, provided such time averages are conducted over a period longer than a characteristic mixing time, they can be replaced by an average over the entire phase space using a measure that is independent of the system's initial microstate;
	that is, they can be replaced by an ensemble average.
	One question that has been raised about this procedure is whether the ensemble average could really be independent of the system's initial conditions in general.
	Consider that any system with a disconnected configuration space (i.e., a system that is not metrically transitive \cite{birkhoff1931proof,neumann1932proof}) is necessarily confined to the component of the configuration space where it begins, meaning that the average behavior of an ensemble of such systems would not resemble the time-averaged behavior of any one system.
	Instead, the measure used for the average should depend on the system's initial microstate, since that defines the component of the configuration space to which the system is confined.
	Following this line of thought further, changes to the connectivity of the configuration space could discontinuously change the support of the measure used for thermodynamic averages, and thereby lead to the discontinuities in thermodynamic observables required for a phase transition.
	This is one way to motivate the topological hypothesis \cite{caiani1997geometry,franzosi2004theorem} which roughly proposes that changes in the topology of the accessible part of the configuration space are necessary for a phase transition to occur.
	
	\begin{figure*}
		\centering
		\includegraphics[width=1.0\textwidth]{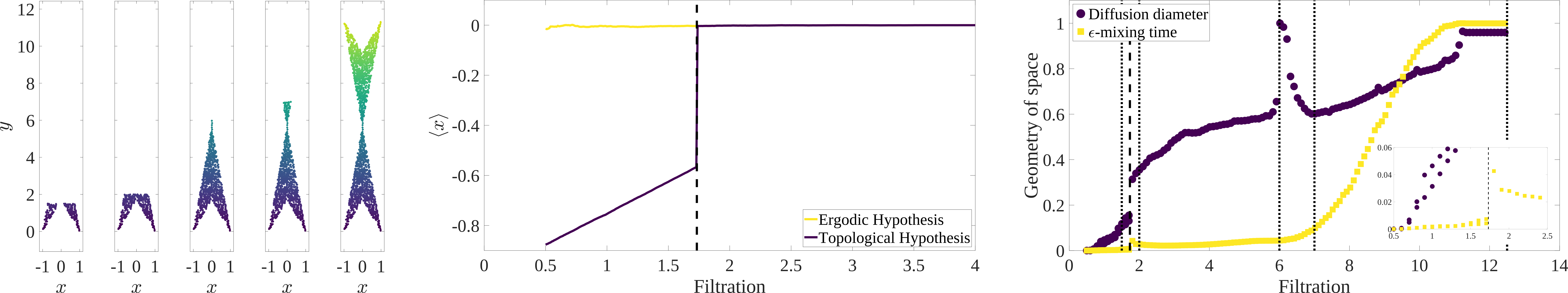}
		\caption{The left panel shows the evolution of a model configuration space with points colored by their $y$-coordinate.
			The filtration value is an upper bound for the $y$-coordinate of any point on the system's trajectory.
			The middle panel shows the average $x$-coordinate $\langle x \rangle$ calculated using the ergodic hypothesis and the topological hypothesis as a function of filtration value, assuming that the system starts at the lower left corner of the configuration space.
			The right panel shows the diffusion diameter and $\epsilon$-mixing time of the configuration space as functions of filtration value and that they are sensitive to changes in the configuration space geometry. 
			The diffusion diameter, but not the mixing time, is also sensitive to the bottleneck.
			The five dotted lines show the filtration values used in the left panel.}
		\label{fig:figure1}
	\end{figure*}
	
	The application of these ideas to a toy model is shown in Fig.\ \ref{fig:figure1}.
	Suppose that there is an isolated system with the configuration space in the left panel, and that the potential energy is strictly monotone increasing with the $y$-coordinate in the figure.
	The system's internal energy then defines a value of $y$ called the filtration value above which the system's trajectory cannot pass.
	Further suppose that the system's initial microstate corresponds to a point in the bottom left corner of the configuration space, and that the observable of interest is the $x$-coordinate in the figure.
	The middle panel shows the ensemble average of this observable as a function of filtration value for both the ergodic hypothesis and the topological hypothesis.
	Since the configuration space is symmetric about $x = 0$ and the ergodic hypothesis stipulates that the measure should be independent of the initial microstate, the ensemble average of the observable is zero for all filtration values.
	By comparison, the topological hypothesis recognizes that for sufficiently small filtration values the configuration space is disconnected, the system is confined to the left side of the configuration space, and the ensemble average of the observable should be zero only when the filtration value increases enough for the space to become connected.
	Moreover, the discontinuous change in the integration measure at this filtration value leads to a discontinuous observable, precisely of the type that would be expected for a system undergoing a phase transition.
	It is significant that it was not necessary at any point to pass to the thermodynamic limit for there to be a discontinuity in the observable, though it is reasonable to suppose that the magnitude of the discontinuities and their location with respect to any control variables would vary systematically with size in a more realistic situation.
	
	Let $E$ be the potential energy of a system, $\vec{q}_i$ be the coordinates of the system's particles, and $V(\vec{q}_1 \dots \vec{q}_N)$ be a smooth, stable, confining and short-range interaction potential.
	The topological hypothesis \cite{franzosi2007topology1,franzosi2007topology2} initially claimed that a topological change in the equipotential energy submanifolds $\Sigma_E = V^{-1}\left((-\infty,E]\right)$ of the configuration space was a necessary condition for a phase transition.
	Kastner and Mehta \cite{kastner2011phase} observed a second-order phase transition in a two-dimensional $\Phi^4$ system at an energy where no topological change occurred, disproving this claim.
	Gori \etal \cite{gori2018topological} subsequently refined the hypothesis and observed that the phase transition in the two-dimensional $\Phi^4$ system was caused by a diverging transition time between two parts of the configuration space, and that this was associated with an asymptotic topological change.
	That is, a continuous change in the configuration space geometry brought about a divergence in the mixing time, and while such events are often accompanied by topological changes they are not always.
	The significance of the configuration space geometry is supported by a recent study of the topological and geometric properties of the two-dimensional $XY$ model by Bel-Hadj-Aissa \etal \cite{bel2021geometrical};
	they computed the mean geometric curvature of the equipotential energy level sets and observed that the location of the phase transition could be inferred from the level set curvatures.
	
	With this as background, the fundamental conjecture of this work is:
	\begin{conjecture}
		\label{conj:phase_transition}
		A necessary condition for a first-order phase transition is a discontinuity in the mixing time on the configuration space.
	\end{conjecture}
	\noindent This is not intended to suggest that the mixing time directly regulates the appearance of phase transitions, but merely that the mixing time is sensitive to any geometric changes that could discontinuously change the integration measure for thermodynamic averages.
	
	A natural question at this point is whether it is actually possible to measure the mixing time of a thermodynamic system, or whether this will always remain a vague quantity significant only in that it is necessarily shorter than the time scales of quasiequilibrium processes.
	We hypothesize that the relevant geometric changes are so severe that any geometric quantity reasonably sensitive to the accessible volume of the configuration space and the length and number of paths connecting distant regions should exhibit measurable discontinuities for the same values of the control variable as the mixing time.
	The two quantities used here are the diffusion distance \cite{coifman2006diffusion,talmon2013diffusion} and the $\epsilon$-mixing time.
	The first measures the difference between two distributions that start as Dirac delta distributions at different locations on the configuration space and evolve by diffusion;
	maximizing the diffusion distance over all starting locations of the two distributions gives the configuration space's diffusion diameter.
	The second measures the time required for a Dirac delta distribution to diffuse to the steady-state distribution within a tolerance defined by $\epsilon$, averaged over all starting locations of the distribution in the configuration space. 
	This is intended to resemble the thermodynamic mixing time, though the distribution evolves by the diffusion equation rather than the Liouville equation and the similarity to the steady-state distribution is quantified by the Kullback--Leibler divergence \cite{kullback1951information}.
	
	The conjecture is tested by evaluating the diffusion diameter and $\epsilon$-mixing time for the configuration spaces of hard disk and hard sphere systems, collectively called hard disk systems in the following.
	These systems are frequently used to model simple fluids \cite{dyre2016simple}, and are governed by the hard disk potential for which the energy is infinite if any disks overlap and is zero otherwise.
	The phase transitions for these systems have been studied extensively, starting with the seminal work of Alder and Wainwright \cite{alder1962phase} who observed a phase transition in a system of hard disks as a function of packing fraction $\eta$ (the fraction of the area covered by the disks).
	The solid for $\eta > 0.72$ is characterized by the presence of a long-range translational and orientational order, whereas the fluid for $\eta < 0.70$ is characterized by the absence of any long-range order \cite{weber1995melting,mitus1997local}.
	The interval $0.70 < \eta < 0.72$ was initially believed to contain coexisting solid and fluid phases as would be expected of a first-order transition, but more recently was claimed to contain a hexatic phase \cite{halperin1978theory,young1979melting}.
	It is significant that there is still not agreement in the literature about the order of the transitions and the phases involved \cite{weber1995melting,mitus1997local,bernard2011two,engel2013hard} even for this simple system.
	Similar studies of hard spheres in three dimensions include those by Isobe and Krauth \cite{isobe2015hard} and Pieprzyk \etal \cite{pieprzyk2019thermodynamic} who observed solid-fluid phase coexistence in the $0.49 < \eta < 0.548$ interval.
	
	The configuration spaces of hard disk systems have been studied before.
	Carlsson \etal \cite{carlsson2012computational} explored the topology of the configuration space of five hard disks in a square box by regularizing the potential energy surface and using classical Morse theory \cite{morse1934calculus,milnor2016morse}.
	Baryshnikov \etal \cite{baryshnikov2014min} developed a min-type Morse theory for the configuration spaces of hard disk systems and proved that the critical points where the topology changes as a function of packing fraction are precisely the mechanically-balanced disk configurations.
	Ritchey \cite{ritcheyphd} studied configuration spaces of hard disk in the hexagonal torus for $n = 1 \ldots 12$ disks, created a database of the critical points, defined their critical index (specifying the nature of the associated topological change), and found their plane symmetry groups.
	The authors previously \cite{ericok2021quotient} proposed distance functions on the configuration spaces of hard disks quotiented by various symmetry groups.
	They subsequently \cite{ericok2021configuration} used these distances to explicitly triangulate the quotients of the configuration space of two spheres in the rhombic dodecahedron and measure the diffusion diameters of the resulting spaces.
	They also observed an accumulation of critical points in the configuration space at packing fractions in the phase coexistence interval, suggesting the possibility of a topological and geometric catastrophe there in the thermodynamic limit.
	
	The technical contribution of this work is a set of techniques described in Sec.\ \ref{sec:configuration_spaces} that collectively allow the configuration space geometries for hard disk systems with $n \leq 7$ disks and hard sphere systems with $n \leq 6$ spheres to be evaluated in practice.
	The diffusion diameters and $\epsilon$-mixing times (defined in Sec.\ \ref{sec:config_space_geometry}) of these spaces are computed as functions of particle diameter and particle number, and in Sec.\ \ref{sec:results} significant discontinuities are observed at the packing fractions of critical points close to the phase coexistence intervals.
	Along with the observation \cite{ericok2021configuration} that critical points accumulate in the phase coexistence intervals with increasing particle number, this suggests that there should be a discontinuity in the mixing time at the packing fraction of the phase transition in the thermodynamic limit, offering preliminary support for Conj.\ \ref{conj:phase_transition}.

	\section{Configuration Spaces}
	\label{sec:configuration_spaces}

	\subsection{Tautological function}
	\label{subsec:tautolocigcal_function}
	
\begin{figure}
		\centering
		\includegraphics[width=1.0\columnwidth]{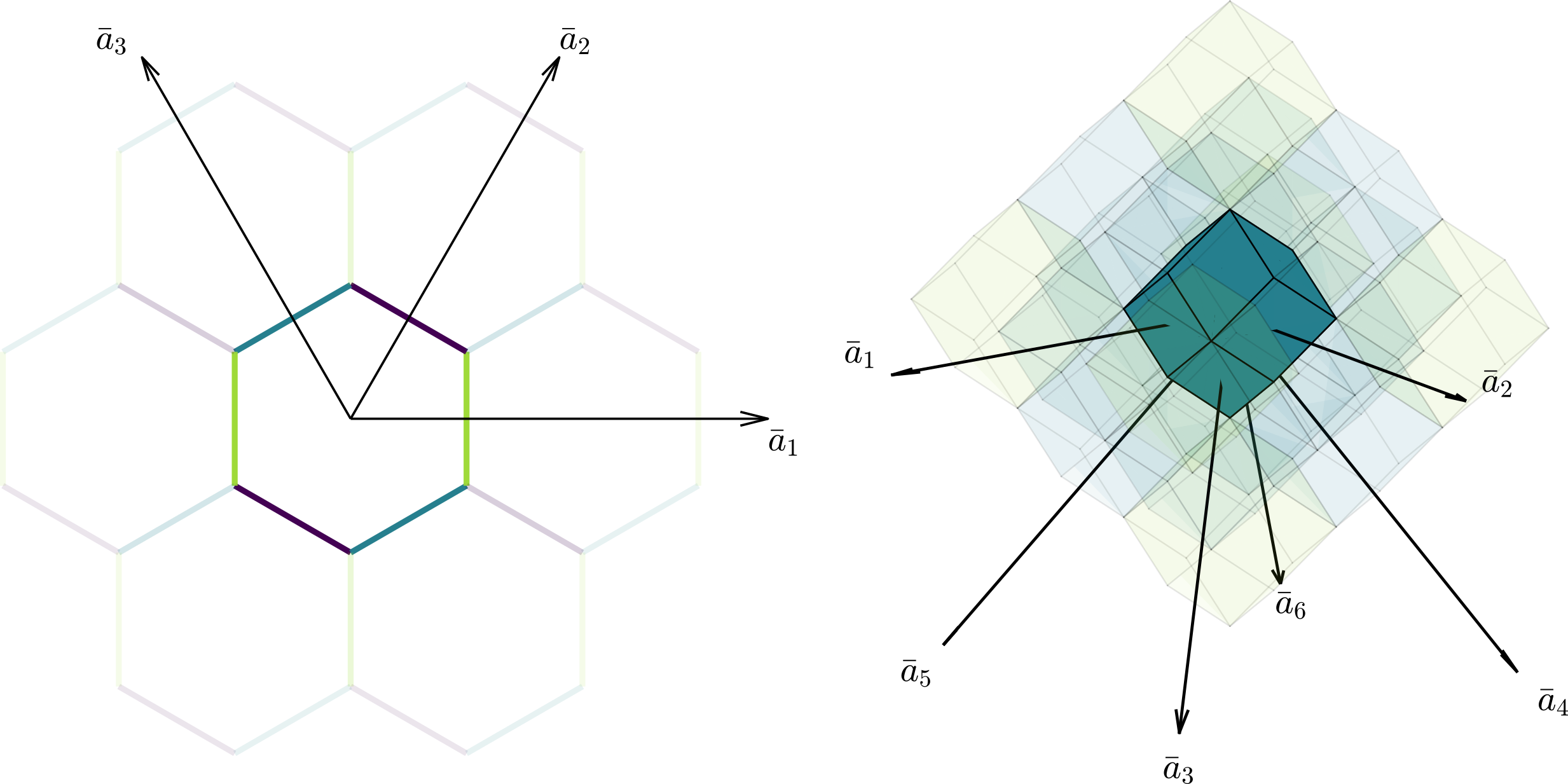}
		\caption{A two-dimensional torus $T^2$ (left) and a three-dimensional torus $T^3$ (right) are shown with some of their periodic images.
			The unit vector $\vec{a}_1$ points in the $x$-direction in both cases.
			All the unit vectors pass through the centers of the corresponding faces.}
		\label{fig:figure2}
	\end{figure}

	The configuration space of $n$ points on a $d$-dimensional torus $T^d$ is
	\begin{equation}
	\Lambda(n) = \{ \config{x} = (\vec{x}_1 \dots \vec{x}_n) \,|\, \vec{x}_i \in T^d \}.
	\end{equation}
	A two-dimensional torus $T^2$ for hard disks is obtained by identifying opposite edges of a regular hexagon, whereas a three-dimensional torus $T^3$ for hard spheres is obtained by identifying opposite faces of a rhombic dodecahedron.
	Figure \ref{fig:figure2} shows these domains with some of the periodic images of the fundamental unit cells.
	The center to center distance to the neighboring cells is one in both cases.
	
	The tautological function $\tau:\Lambda(n) \rightarrow R$ is the maximum radius that the disks could have without overlapping, or
	\begin{equation}
	\tau(\config{x}) = \min_{\substack{1 \le i < j \le n}} r_{ij}
	\label{eq:tautological_function}
	\end{equation}
	where $r_{ij}$ is half the geodesic distance between the centers of disks $i$ and $j$. This allows the configuration space of $n$ disks of radius $\rho$ to be written as
	\begin{equation}
	\Gamma(n,\rho) = \tau^{-1}\left([\rho,\infty)\right)
	\label{eq:config_space_full}
	\end{equation}
	or the set of all configurations of points where the minimum point separation is at least $\rho$.

	\subsection{Critical configurations}
	\label{subsec:critical_configurations}
	
	Morse theory \cite{morse1934calculus,milnor2016morse} stipulates that the topology of the sublevel sets of a generic real-valued function $f$ defined on a smooth manifold $M$ can change only at the critical points of $f$.
	Intuitively, a critical point is a point where the gradient of the function vanishes.
	The index of a critical point is defined as the number of negative eigenvalues of the Hessian matrix there, and roughly indicates the number of independent directions along which $f$ decreases to second order.
	Let $M_a = \{x \in M | f(x) < a\}$ denote a sublevel set of $M$.
	One of the results of Morse theory is that the topology of $M_a$ and $M_b$ are equivalent if the interval $[a, b]$ doesn't contain any critical points of $f$.
	
	Baryshnikov \etal \cite{baryshnikov2014min} began to develop a min-type Morse theory specifically for hard disk configuration spaces, and proved that the critical points of the tautological function are precisely the mechanically-balanced configurations.
	These can be shown to coincide with the critical points of the regularized hard-disk potential in Ref.\ \cite{ericok2021configuration}, found using a conjugate gradient algorithm to minimize the square of the potential gradient magnitude.
	Interactive databases of the critical points for the hard disk\footnote{\url{https://github.com/burakericok/hard_disk_crits}} and hard sphere\footnote{\url{https://github.com/burakericok/hard_sphere_crits}} configuration spaces are available online.
	The distributions of critical points as functions of index and packing fraction in Refs.\ \cite{ritcheyphd,ericok2021configuration} suggest that an accumulation of critical points in a narrow packing fraction interval with increasing number of disks could be associated with the onset of a phase transition;
	one of the purposes of this work is to explore this observation further.

	\subsection{Geometric representation}
	\label{subsec:graph_network}
	
	The procedure followed previously \cite{ericok2021quotient,ericok2021configuration} when studying hard disk configuration spaces involved repeatedly sampling points in the configuration space of points $\Lambda(n)$, connecting nearby points in the resulting point cloud to reconstruct $\Lambda(n)$, and restricting to the hard disk configuration space $\Gamma(n, \rho)$ by retaining only those regions with suitable values of the tautological function.
	The difficulty with this procedure is that the dimension and complexity of $\Gamma(n, \rho)$ increases rapidly with $n$, quickly requiring a point cloud with an overwhelming number of points to accurately capture the geometric details.
	This section describes several techniques that, while they do not solve the problem, reduce the required number of points enough to allow us to study the geometry of configuration spaces with up to $18$ dimensions.
	
	The first is to quotient the configuration space by various symmetry groups, thereby reducing the relevant volume.
	The three isometry groups considered here are the group of rigid translations $\mathcal{T}$, the group of disk label permutations $\mathcal{P}$, and the point group symmetries of the fundamental unit cell $\mathcal{L}$.
	As described previously \cite{ericok2021quotient,ericok2021configuration}, the configuration space is quotiented by these groups by changing the distance function used to identify and connect nearby points in the point cloud.
	More precisely, there is a natural distance function $d_{\Lambda}(\config{x}, \config{y}) = \sum_{i = 1}^{n} \lVert \vec{x}_i - \vec{y}_i \rVert$ on $\Lambda$ that is the sum of the geodesic distances each point in a configuration $\config{x}$ would need to travel to be converted into configuration $\config{y}$.
	Given an isometry group $\mathcal{S}$, the distance $d_{\Lambda/\mathcal{S}}$ in the quotient space $\Lambda/\mathcal{S}$ can be written as \cite{burago2001course}
	\begin{equation}
	d_{\Lambda/\mathcal{S}}(\config{x},\config{y}) = \inf \limits_{\substack{S \in \mathcal{S}}} d_\Lambda [\config{x},S(\config{y})].
	\label{eq:distance_quotient}
	\end{equation}
	Observe that $\mathcal{T}$ is a continuous group, that Eq.\ \ref{eq:distance_quotient} involves solving a global optimization problem when $\mathcal{T} \subset \mathcal{S}$, and that this problem can have multiple local minima.
	While algorithms like Tabu search \cite{glover1989tabu,chelouah2000tabu} can be used in such situations, the number of problems to be solved increases rapidly with the number of isometry groups considered.
	For example, when $\mathcal{S} = \mathcal{T} \cup \mathcal{P} \cup \mathcal{L}$, the global optimization over rigid translations needs to be solved $n! \times O(\mathcal{L})$ times to evaluate Eq.\ \ref{eq:distance_quotient} where $O(\mathcal{L})$ is the order of $\mathcal{L}$.
	
	The second technique reduces the number of optimization problems that need to be solved in Eq.\ \ref{eq:distance_quotient}.
	The idea is to construct an approximation to the solution of the global optimization problem over rigid translations, allowing the discrete symmetry operations that are unlikely to realize the minimum to be quickly rejected.
	The procedure followed here samples a fixed number of random translations for a fixed discrete symmetry operation, and uses the minimum over the random translations as the approximation.
	The details of this procedure and various numerical results are provided in App.\ \ref{sec:heuristic}.
	This reduces the computational cost enough to be able to evaluate a small number of distances $d_{\Lambda/\mathcal{S}}$, but not all the pairwise distances in the point cloud when identifying nearby points to reconstruct $\Lambda/\mathcal{S}$.
	
	The third techniques reduces the number of times that $d_{\Lambda/\mathcal{S}}$ needs to be evaluated.
	Points sampled from $\Lambda$ are initially mapped to a Cartesian space with coordinates given by descriptors that are invariant to the action of $\mathcal{T}$, $\mathcal{P}$, and $\mathcal{L}$ \cite{ericok2021quotient,ericok2021configuration}.
	This mapping is conjectured to be an embedding and therefore to preserve the neighborhood of every point in $\Lambda$, but is not isometric in the sense that distances between configurations are distorted.
	That said, there are efficient algorithms to calculate $k'$-nearest neighbor graphs in the descriptor space with the Euclidean metric.
	Since neighborhoods are preserved, the $k'$-nearest neighbor graph of a point in the descriptor space should contain the $k$-nearest neighbor graph of the point in $\Lambda/\mathcal{S}$ for sufficiently large $k' > k$;
	our numerical experiments suggest that $k' \sim 5k$ is usually sufficient. 
	The procedure therefore involves constructing the $k'$-nearest neighbor graph in the descriptor space, evaluating $d_{\Lambda/\mathcal{S}}$ for each edge of this graph, and constructing the approximate $k$-nearest neighbor graph in $\Lambda/\mathcal{S}$ using these distances.
	
\begin{figure}
		\centering
		\includegraphics[width=1.0\columnwidth]{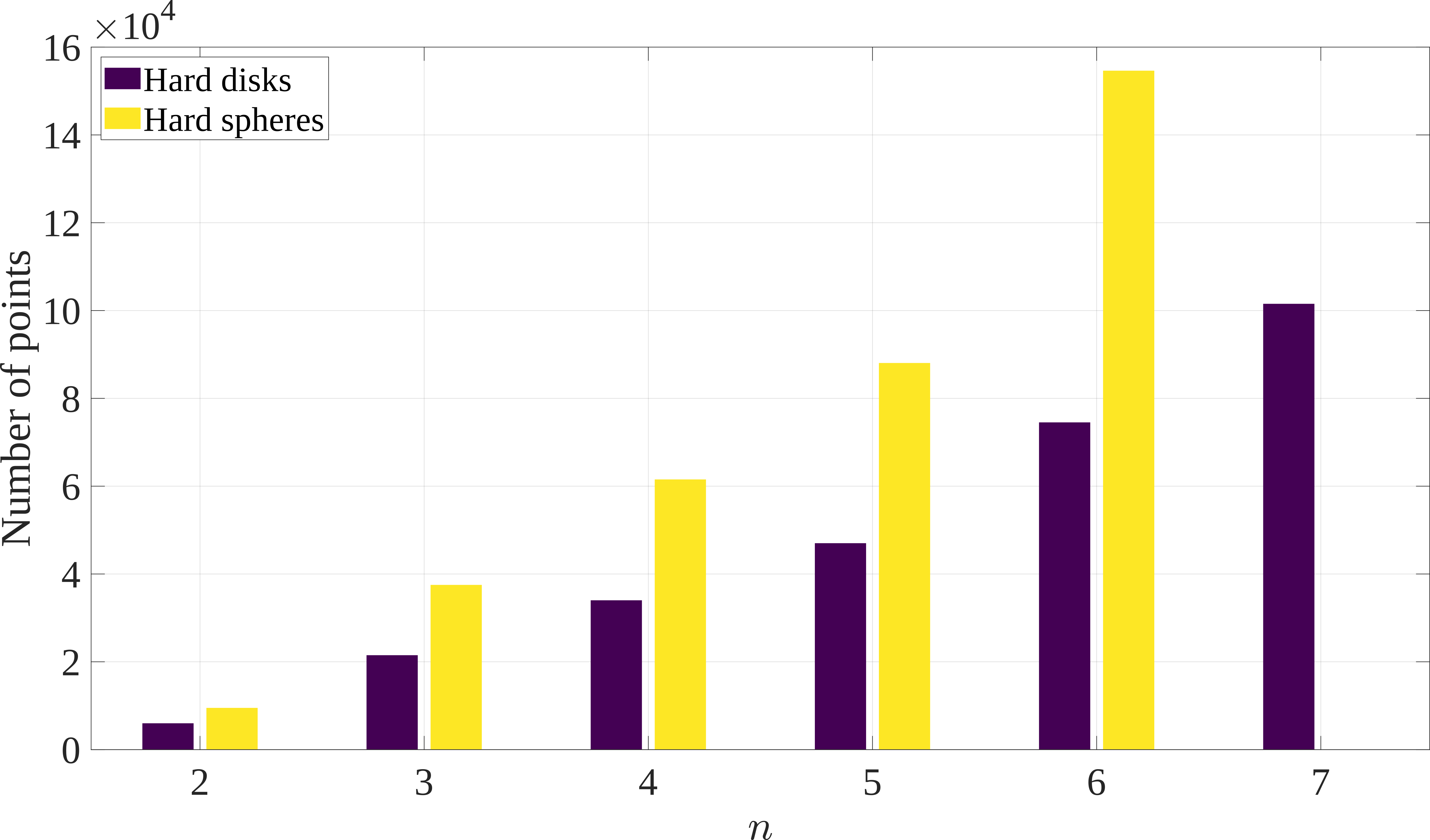}
		\caption{Number of points used in the graph representations for hard disks and spheres.}
		\label{fig:figure3}
	\end{figure}

	It is useful at this point to discuss the relationship of the $k$-nearest neighbor graph in $\Lambda/\mathcal{S}$ to the geometry of the underlying space.
	It is clear that the apparent connectivity of the space depends on the value of $k$;
	for very small $k$ the graph would likely contain many disconnected components, whereas for very large $k$ every point would share an edge with every other point, regardless of the actual properties of the space.
	While there does not seem to be an established canonical approach to selecting the value of $k$, the strategy followed here involves the use of topological information.
	Observe that, for any filtration of the $k$-nearest neighbor graph by the tautological function, the number of disconnected components in the resulting graph should be at most the number of index-$0$ critical points with disk radius larger than or equal to the filtration value.
	Moreover, when the filtration value is equal to the smallest value such that all of the index-$1$ critical points (which correspond to saddle points) are included, only a single connected component should remain.
	The smallest value of $k$ for which these conditions are satisfied seems a reasonable choice, and for the numbers of points in our point clouds (indicated in Fig.\ \ref{fig:figure3}) is approximately $k = 100 \times n$ where $n$ is the number of disks.
	
\begin{figure}
		\centering
		\includegraphics[width=1.0\columnwidth]{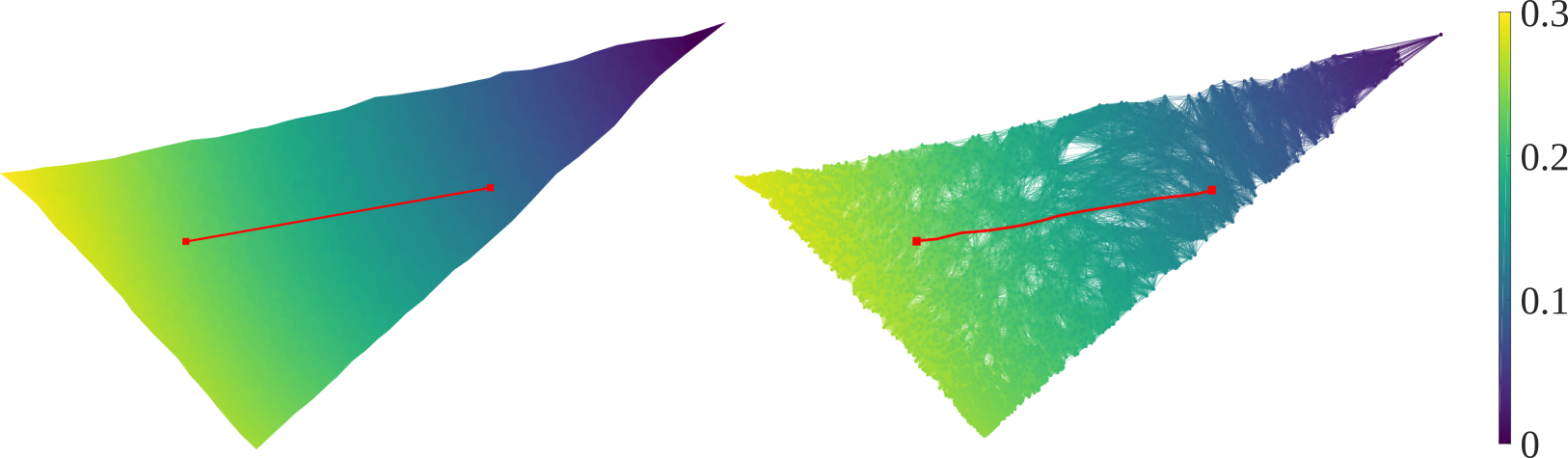}
		\caption{The quotient space $\Lambda(2)/\{\mathcal{T,P,L}\}$ of two points on $T^2$ represented as an $\alpha$-complex (left) and a graph (right), with color indicating the value of the tautological function. The red path connecting two points is a straight line on the left, but is slightly longer on the right.}
		\label{fig:figure4}
	\end{figure}

	The fourth technique is to directly use the $k$-nearest neighbor graph rather than a simplicial complex (a triangulation) to evaluate the geometric properties of $\Lambda/\mathcal{S}$.
	Previously, the configuration spaces for small numbers of hard disks and spheres were triangulated as $\alpha$-complexes \cite{ericok2021quotient,ericok2021configuration}.
	While such simplicial complexes can accurately represent all the geometric properties of the underlying space, the number of simplices required generally increases exponentially with dimension, quickly making the computational memory requirements prohibitive \cite{sheehy2013linear,buchet2016efficient}.
	Fortunately, the only geometric information required to calculate the diffusion diameter and the $\epsilon$-mixing time are geodesic distances, and these can be reasonably approximated from the $k$-nearest neighbor graph alone.
	Figure \ref{fig:figure4} shows the quotient space $\Lambda(2)/\{\mathcal{T,P,L}\}$ of two points on $T^2$ represented as an $\alpha$-complex (left) and as a graph (right).
	While the shortest path connecting two points is a straight line on the left, the path through the edges of the graph on the right is slightly longer.
	Since both the diffusion distance and the $\epsilon$-mixing time are designed to be robust to small geometric perturbations like these, using the $k$-nearest neighbor graph of $\Lambda/\mathcal{S}$ directly is sufficient for our purposes.
	
\begin{figure*}
		\centering
		\includegraphics[width=1.0\textwidth]{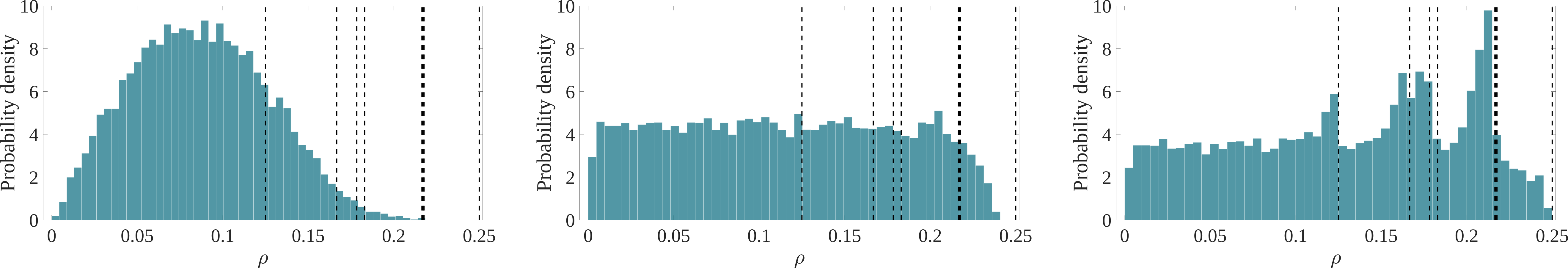}
		\caption{Distributions of tautological function values for point clouds on $\Lambda(4)$ sampled by three different procedures, with the dashed lines indicating the radii of previously-identified critical points.
			Sampling points uniformly gives the distribution on the left.
			The distribution in the middle uses importance sampling and is more uniform.
			The distribution on the right is for a point cloud that both uses importance sampling and samples additional points in the neighborhoods of the critical points.}
		\label{fig:figure5}
	\end{figure*}

	The fifth and final technique is related to the construction of the point cloud on $\Lambda$, and is intended to reduce the number of points required to accurately represent the relevant geometric features of $\Lambda/\mathcal{S}$.
	Figure \ref{fig:figure5} shows the distributions of tautological function values $\rho$ for point clouds on $\Lambda(4)$ sampled by three different procedures, with the dashed lines indicating the radii of previously-identified critical points.
	The left panel shows the distribution of points sampled uniformly on $\Lambda(4)$, and reveals that the overwhelming majority of the space's volume has comparatively small values of $\rho$ where only a single pair of disks would be in contact.
	The critical points are concentrated at relatively high values of $\rho$ though.
	This motivates the use of importance sampling \cite{kahn1951estimation} to sample points more uniformly over $\rho$, concentrating points in the regions where the topology and geometry of $\Gamma(4, \rho)$ are most likely to change and giving the distribution in the middle panel.
	The sampling density around the critical points can be increased further by perturbing known critical configurations and adding the resulting configurations directly to the point cloud, giving the distribution in the right panel.
	The intention is to ensure that the density of sampled points is high enough to accurately reflect the parts of $\Lambda$ that are most relevant to the hypothesis without unnecessarily sampling points elsewhere in the space.
	
	\section{Configuration space geometry}
	\label{sec:config_space_geometry}
	
	Diffusion is a smoothing process.
	Flows from regions of high concentration to low concentration necessarily make a concentration field more uniform with time.
	This also makes diffusion processes relatively insensitive to small perturbations in the initial concentration field or to the geometry of the underlying space.
	This property is likely one of the reasons that diffusive processes are often used to learn about the geometry of a space represented by a sampled point cloud, since the resulting insights do not depend sensitively on the distribution of the points.

	\subsection{Diffusion Distance}
	\label{subsec:diffusion_distance}
	
	Let $G$ be the $k$-nearest neighbor graph on $\Lambda/\mathcal{S}$ constructed by the procedure in Sec.\ \ref{subsec:graph_network}.
	The diffusion distance $d_{ij, t}$ on $G$ measures the $L^2$ distance between two distributions that begin as Dirac delta distributions on vertices $i$ and $j$ and diffuse on $G$ for a time $t$.
	Let the kernel matrix $\mat{K}$ have entries $k_{ij} = \exp(-l_{ij}^2/\sigma^2)$ where $l_{ij}$ is Dijkstra's distance between vertices $i$ and $j$ and $\sigma$ is twice the median length of edges in $G$, and the diagonal degree matrix $\mat{D}$ has entries $d_{ii} = \sum_j k_{ij}$.
	Then $\mat{P} = \mat{D}^{-1} \mat{K}$ is the transition rate matrix of a continuous-time Markov process where the effect of raising $\mat{P}$ to the $t$th power is equivalent to propagating the process by a time $t$.
	Let $\{\lambda_l, \vec{\phi}_l\}$ be the set of eigenvalues and eigenvectors of $\mat{P}$ for $0 \le l \le N - 1$ where $N$ is the number of vertices.
	Since $\mat{P}$ is normalized, the largest eigenvalue $\lambda_0$ is associated with a constant eigenvector $\vec{\phi}_0$.
	Discarding $\lambda_0$ and $\vec{\phi}_0$, the diffusion coordinates are defined as
	\begin{equation}
	\vec{\Phi}_{t,i} = \left[\lambda_1^t \vec{\phi}_1(i), \dots, \lambda_{N - 1}^t \vec{\phi}_{N - 1}(i)\right]
	\label{eq:diffusion_maps}
	\end{equation}
	and encode a distribution starting as a Dirac delta distribution on vertex $i$ and diffusing for a time $t$.
	The diffusion distance between vertices $i$ and $j$ is then defined as
	\begin{equation}
	d_{ij,t} = \norm{ \vec{\Phi}_{t,i} - \vec{\Phi}_{t,j} }
	\label{eq:diffusion_distance}
	\end{equation}
	where $\norm{\cdot}$ is the Euclidean norm.
	Note that the significance of the time $t$ depends on the diffusion rate implicit in the kernel $\mat{K}$. Since the eigenspectrum of $\mat{P}^t$ decays very quickly for sufficiently large $t$, the diffusion distance $d_{ij,t}$ can often be accurately approximated using only the first few eigenvalues and eigenvectors for a substantial time and memory savings \cite{talmon2013diffusion}.
	Details about this approximation are provided in App.\ \ref{sec:eigenspectrum_analysis}.

	\subsection{Mixing Time}
	\label{subsec:mixing_time}
	
	Thermodynamics is generally concerned with equilibrium or quasiequilibrium systems, where there is no net redistribution of matter or energy and thermodynamic observables are time independent.
	From the standpoint of statistical thermodynamics, this means that the time average of any thermodynamic observable over the system's trajectory through the phase space does not depend on the initial microstate provided that the averaging is performed over a sufficiently long time interval; the shortest such interval is known as the mixing time.
	
	Let $\Omega$ be the phase space of a thermodynamic system, $\vec{q}$ and $\vec{p}$ be the canonical coordinates and momenta, and $\mu(\vec{q}, \vec{p}, t)$ be the probability distribution of microstates on $\Omega$ at time $t$.
	Suppose that the system has an arbitrary initial condition $\mu_0(\vec{q}, \vec{p}) = \mu(\vec{q}, \vec{p}, 0)$.
	Briefly setting aside the specifics of the time evolution of $\mu(\vec{q}, \vec{p}, t)$, the steady-state distribution $\mu_\infty(\vec{q}, \vec{p}) = \lim_{t \rightarrow \infty} \mu(\vec{q}, \vec{p}, t)$ defines the equilibrium condition on $\Omega$, and the mixing time could in principle be defined by the approach of $\mu(\vec{q}, \vec{p}, t)$ to $\mu_\infty(\vec{q}, \vec{p})$.
	That said, the literature does not seem to precisely define the necessary conditions for the two distributions to be effectively indistinguishable, nor the effect of the initial condition $\mu_0(\vec{q}, \vec{p})$ on the resulting mixing time.
	A likely reason for this is the historical emphasis on equilibrium states for which, by hypothesis, the observation time can be made as long as necessary for there to be complete mixing.
	
	Classically, the time evolution of $\mu(\vec{q}, \vec{p}, t)$ is governed by the Liouville equation \cite{liouville1838note,gibbs1902elementary}.
	Liouville's theorem states that the resulting convective derivative of the probability distribution is zero, or that the flow of probability density resembles that of an incompressible fluid.
	This raises questions relating to the convergence of probability distributions that are closely related to those about the origins of irreversibility, and that continue to be discussed in the literature.
	This discussion is avoided by simply supposing that $\mu(\vec{q}, \vec{p}, t)$ evolves by the diffusion equation, while not endorsing this approach generally.
	One consequence of this supposition is that the marginal probability distributions on the configuration and momentum subspaces evolve independently, though only the marginal probability distribution on the configuration space has a limiting distribution (the momentum subspace is unbounded).
	This suggests that the mixing time be defined by the convergence of the marginal probability distribution $\nu(\vec{q}, t)$ on the configuration space $\Omega_q$ to the limiting distribution $\nu_\infty(\vec{q})$.
	
	The Kullback-Leibler divergence \cite{kullback1951information} is often called the relative entropy, and is a standard way to quantify how much a probability distribution $\nu$ differs from a reference probability distribution $\nu_\infty$:
	\begin{equation}
	I(t; \nu_0) = \int_\Omega \nu(\vec{q}, t) \log \frac{\nu(\vec{q}, t)}{\nu_\infty(\vec{q})} \diff \vec{q}.
	\label{eq:kl_div}
	\end{equation}
	We propose that the condition for $\nu(\vec{q}, t)$ to have converged to $\nu_\infty(\vec{q})$ be that $I(t; \nu_0) \leq \epsilon$, where $\epsilon$ is an adjustable parameter analogous to a conventional convergence threshold.
	
	Now suppose that $\nu_0$ is a Dirac delta distribution centered at $\vec{q}_0$, and that $t_\epsilon(\vec{q}_0)$ is the minimum time required for this initial condition to converge to $\nu_\infty(\vec{q})$ on $\Omega_q$ in the sense above.
	The $\epsilon$-mixing time $\langle t \rangle_\epsilon$ is defined as the weighted average of $t_\epsilon$ over all possible choices of the initial condition, or
	\begin{equation}
	\langle t \rangle _\epsilon = \int_\Omega t_\epsilon(\vec{q}) \nu_\infty(\vec{q}) d\vec{q}
	\label{eq:average_epsilon_mixing_time}
	\end{equation}
	where the measure of integration is taken to be $\nu_\infty(\vec{q})$ in the absence of a natural alternative. 
	The $\epsilon$-mixing time is regarded as a precisely-defined proxy for the thermodynamic mixing time on the configuration space in the following.
	
	This leaves only the definition of a diffusive process occurring on a connected graph $G$ instead of on a continuous space $\Omega_q$.
	Let $\vec{\nu}(t)$ be the probability masses on the vertices of $G$ at time $t$, and define the graph Laplacian as $\mat{L} = \mat{D} - \mat{K}$ \cite{chung1997spectral}.
	The governing equation for a diffusive process is
	\begin{equation}
	(\partial/\partial t + \mat{L}) \vec{\nu}(t) = 0. 
	\label{eq:diffusion_equation}
	\end{equation}
	Let $\mat{\Lambda}$ be the diagonal matrix of the eigenvalues $\lambda_0 = 0 < \lambda_1 \dots \le \lambda_{N-1}$ of $\mat{L}$, and $\mat{\Phi}$ be the matrix of the corresponding eigenvectors where the first column is a constant vector.
	Since $\mat{L}$ is a symmetric matrix, the columns of $\mat{\Phi}$ form an orthogonal basis and the solution to the diffusion equation is
	\begin{equation}
	\vec{\nu}(t) = \mat{\Phi} e^{-t\mat{\Lambda}} \mat{\Phi}^{T} \vec{\nu}_0
	\label{eq:transient_concentation}
	\end{equation}
	for the initial condition $\vec{\nu}_0$.
	The steady-state distribution is readily calculated using the fact that all of the diagonal elements of $e^{-t\mat{\Lambda}}$ go to zero in the limit of long time except the first term which goes to one.
	This suggests that the steady-state distribution $\nu_\infty$ is always the uniform distribution over all vertices of $G$.
	
	The right panel of Fig.\ \ref{fig:figure1} shows the different behaviors of the diffusion distance and the mixing time in practice.
	Observe that while both exhibit a discontinuous jump at the filtration value where the two components of the space merge, the diffusion diameter has an additional peak at the filtration value of the bottleneck.
	% \blue{In the case of the bottleneck, the diffusion diameter is the diffusion distance between the bottleneck and one of the two corner points (equally likely because of the symmetry). Since the diffusion distance considers all possible paths connecting pair of points, this can be interpreted as having lesser number of paths connecting the bottleneck to the other corners and a larger diffusion distance as a result. However, when the bottleneck is passed, the diffusion diameter becomes the diffusion distance between the two points that are diagonally the most distant from each other (\ie bottom left and top right corners or bottom right and top left corners). In this case, more paths connecting these points appear and the diffusion distance goes down compared to the bottleneck. The overall increasing trend of the diffusion diameter is due to the growth of the space. }  
	% This is the result of the volume of the bottleneck being very small compared to the rest of the space.
	The initial rise is attributed to the probability mass having difficulty diffusing to the point of the bottleneck through a small number of paths.
	Since the space is constructed by means of $k$-nearest neighbor graphs, more paths spanning the neck appear with further increases in the filtration value, subsequently reducing the diffusion diameter.
	That is, the peak in the diffusion diameter is effectively an artifact of the construction of the space and the precise sequence in which the edges appear.
	By comparison, the average mixing time $\langle t \rangle _\epsilon$ is unaffected by the bottleneck since the small volume of that region reduces its contribution to the average in Eq.\ \ref{eq:average_epsilon_mixing_time}.
	More details about the calculation of the $\epsilon$-mixing time for this system are given in App.\ \ref{sec:mixing_time_toy}.

	\section{Results}
	\label{sec:results}
	
	The hard disk equation of state usually appears in the literature as a function of packing fraction $\eta$ \cite{mulero2008equations,tian2020equations}.
	The radius $\rho$ of the disks appearing in Eq.\ \ref{eq:config_space_full} can be converted to a packing fraction using $\eta = n \pi \rho^2 / A$ for hard disks and $\eta = 4 n \pi \rho^3 / (3 V)$ for hard spheres, where $A = \sqrt{3}/2$ and $V = \sqrt{2} / 2$ are respectively the area of the fundamental hexagon and the volume of the fundamental rhombic dodecahedron in Fig.\ \ref{fig:figure2}.
	
	The authors previously constructed quotient spaces of two hard disks \cite{ericok2021quotient} and two hard spheres \cite{ericok2021configuration} and studied the topological and geometric properties of both the original and quotient spaces as functions of packing fraction.
	They observed that while the geometric and topological features of the these spaces changed dramatically with the choice of symmetry groups by which to quotient, the general behavior of the diffusion diameter (and significantly the locations of the discontinuities) as a function of packing fraction did not.
	This suggests that the quotient space has the advantage of a substantially reduced volume while retaining the essential geometric and topological features of the full configuration space.
	If the behavior of the $\epsilon$-mixing time is similarly robust to the choice of symmetry groups by which to quotient, this would allow Conj.\ \ref{conj:phase_transition} to be tested more easily and for systems with larger numbers of disks.
	
\begin{figure}
		\centering
		\includegraphics[width=1.0\columnwidth]{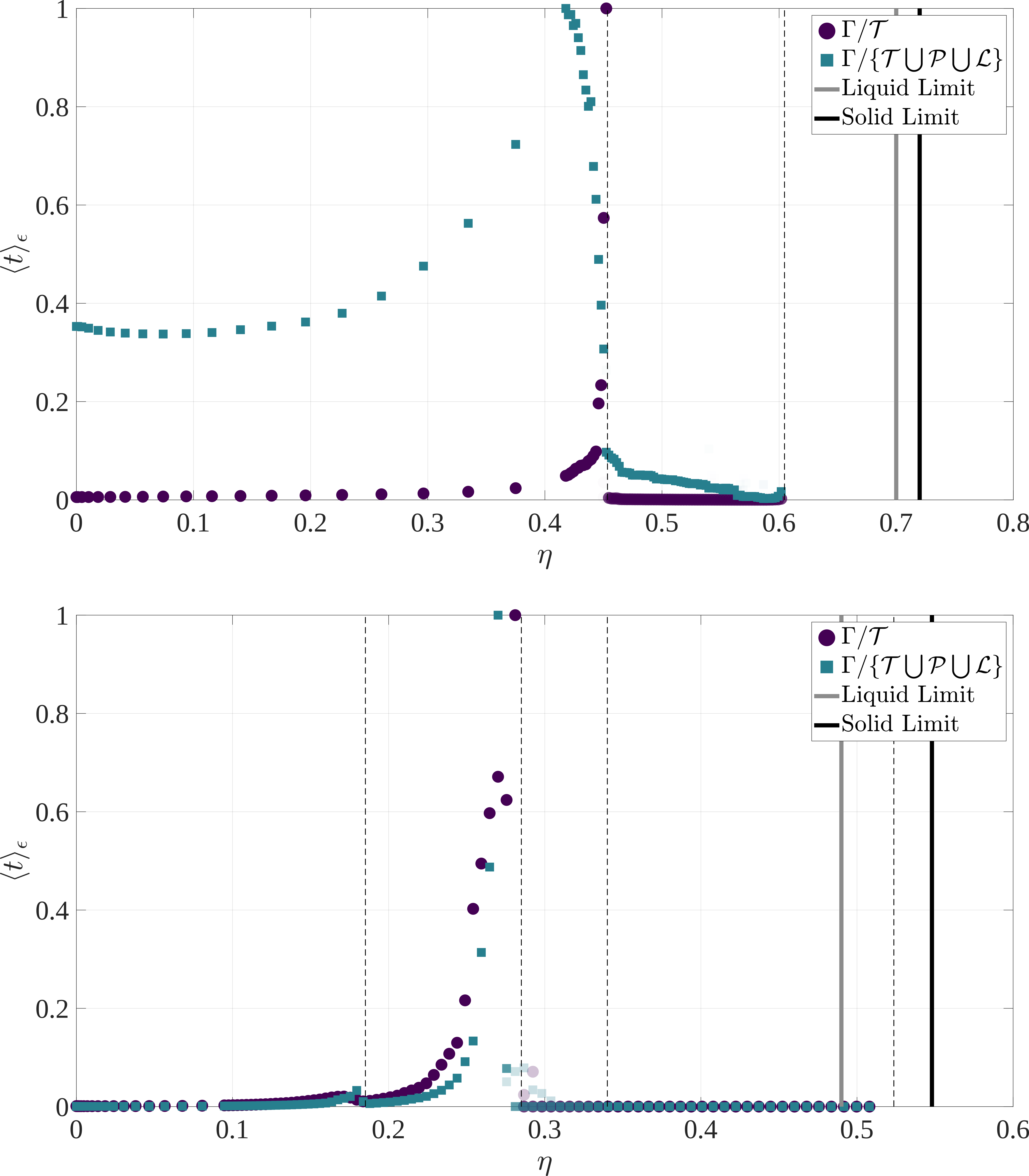}
		\caption{$\epsilon$-mixing times of the quotient spaces $\Gamma(2,\eta)/\mathcal{T}$ and $\Gamma(2,\eta)/\{\mathcal{T} \cup \mathcal{P} \cup \mathcal{L}\}$ for hard disks (top) and hard spheres (bottom).
			Dashed lines represent the packing fractions of the critical points.}
		\label{fig:figure6}
	\end{figure}

	Figure \ref{fig:figure6} shows the $\epsilon$-mixing times of the quotient spaces $\Gamma(2,\eta)/\mathcal{T}$ and $\Gamma(2,\eta)/\{\mathcal{T} \cup \mathcal{P} \cup \mathcal{L}\}$ for hard disks (top) and hard spheres (bottom).
	Since only relative changes in the $\epsilon$-mixing time are significant, all the results in this section are normalized to their maximum value for ease of comparison.
	Initially consider the hard disk results in the top row.
	There are only two critical points, one index-$0$ and one index-$1$.
	The volumes and $\epsilon$-mixing times of both spaces initially grow slowly with decreasing $\eta$, with a discontinuity appearing at precisely the packing fraction of the index-$1$ critical point;
	the discontinuity is stronger for $\Gamma(2,\eta)/\mathcal{T}$ since the geometric change is more severe, though at the price of greatly increased computational cost.
	The bottom row shows the results for hard spheres, for which there are two distinct index-$0$, one index-$1$, and one index-$2$ critical points.
	Since having distinct index-$0$ critical points results in the space having multiple disconnected components for certain intervals of $\eta$, the $\epsilon$-mixing times for each component are shown with an opacity that indicates the fraction of vertices participating in that component.
	As before, there is a discontinuity in the $\epsilon$-mixing time at the packing fraction of the index-$1$ critical point, in this situation indicating that the disconnected components are joined.
	Notice that a similar discontinuity does not occur at the packing fraction of the index-$2$ critical point, nor indeed at any other critical point of the tautological function.
	Since the behavior of the diffusion diameter was previously found to be similarly robust to the choice of symmetry groups by which to quotient, the quotient spaces $\Gamma/\{\mathcal{T} \cup \mathcal{P} \cup \mathcal{L}\}$ will be used exclusively in the following.
	
\begin{figure*}
		\centering
		\includegraphics[width=1.0\textwidth]{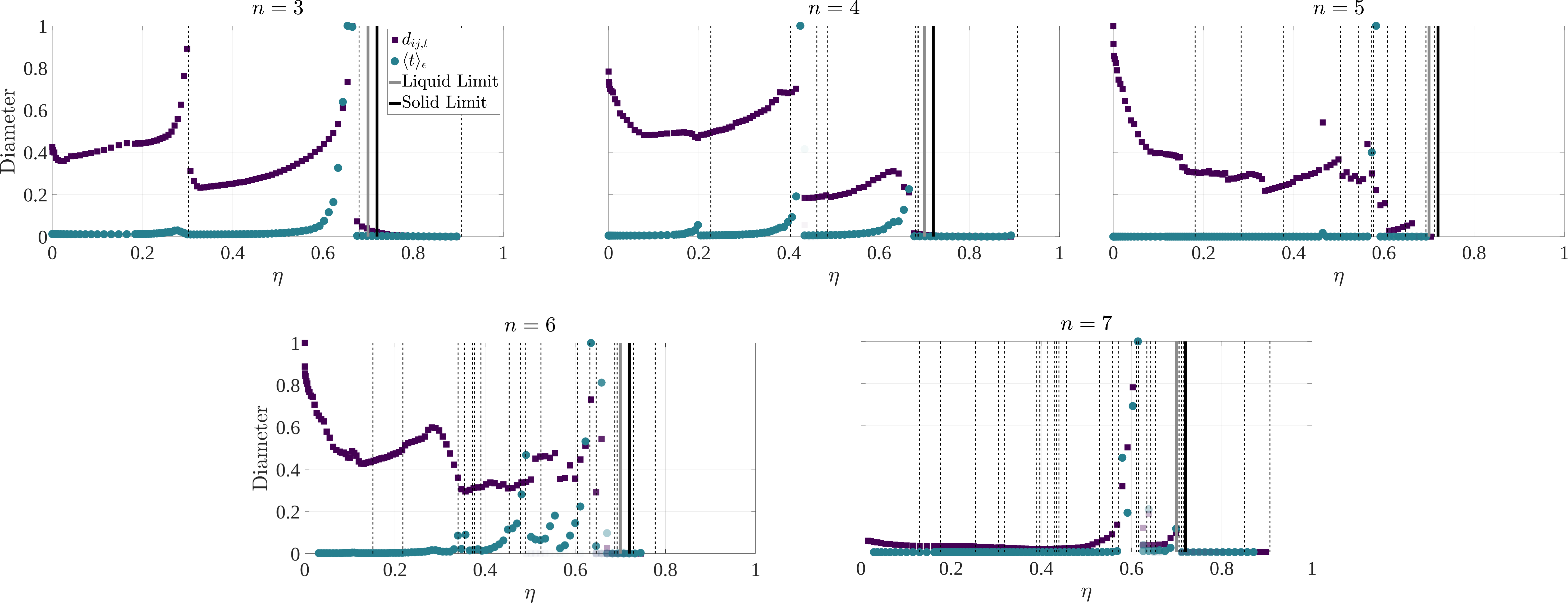}
		\caption{Diffusion diameters and $\epsilon$-mixing times of the spaces $\Gamma/\{\mathcal{T} \cup \mathcal{P} \cup \mathcal{L}\}$ for $n = 3 \dots 7$ hard disks as functions of packing fraction.
			Dashed lines shows the packing fractions of the critical points.}
		\label{fig:figure7}
	\end{figure*}

	Figure \ref{fig:figure7} shows the diffusion diameters and $\epsilon$-mixing times of the quotient spaces $\Gamma/\{\mathcal{T} \cup \mathcal{P} \cup \mathcal{L}\}$ for $n = 3 \dots 7$ hard disks as functions of packing fraction, with dashed lines indicating the packing fractions of the critical points.
	The first observation is that not all of the critical points correspond to substantial geometric changes, at least not ones to which the diffusion distance and $\epsilon$-mixing time are sensitive.
	The second is that the diffusion diameter is generally much noisier than the $\epsilon$-mixing time, and while the structure in the signal could perhaps be analyzed for further geometric information that is not the purpose of our study.
	The third is that while the discontinuities in the diffusion distance and $\epsilon$-mixing time do not always occur at the same packing fractions, the largest discontinuities generally do.
	
\begin{figure*}
		\centering
		\includegraphics[width=0.7\textwidth]{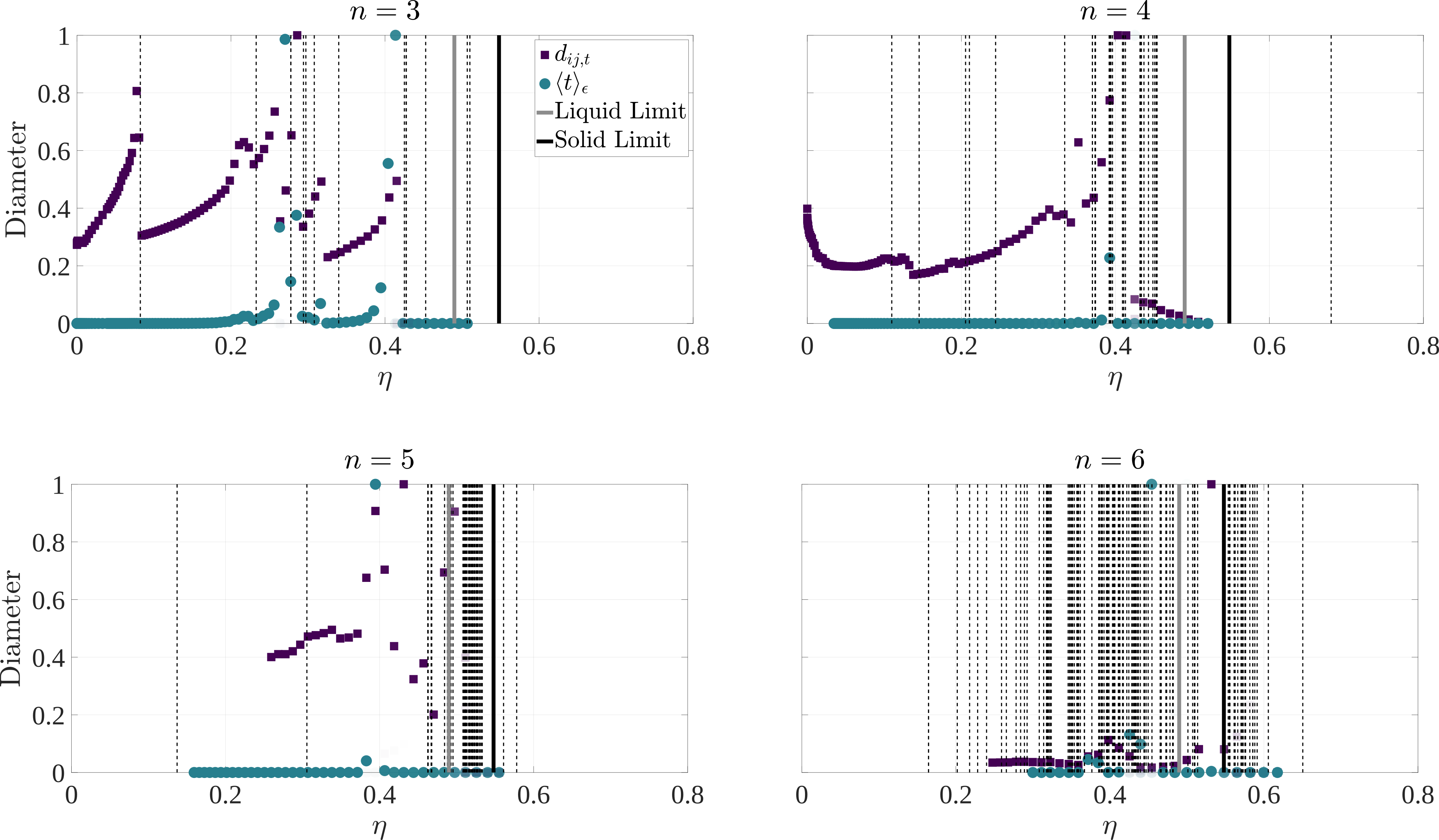}
		\caption{Diffusion diameters and $\epsilon$-mixing times of the spaces $\Gamma/\{\mathcal{T} \cup \mathcal{P} \cup \mathcal{L}\}$ for $n = 3 \dots 6$ hard spheres as functions of packing fraction.
			Dashed lines shows the packing fractions of the critical points.}
		\label{fig:figure8}
	\end{figure*}

	Figure \ref{fig:figure8} shows the corresponding results to Fig.\ \ref{fig:figure7} for $n = 3 \dots 6$ hard spheres.
	While the number of critical points is greatly increased relative to the hard disk systems in Fig.\ \ref{fig:figure7}, the number of discontinuities in the diffusion diameter and $\epsilon$-mixing time are approximately the same.
	This suggests that there is perhaps a small class of critical points associated with substantial geometric changes to the configuration space, and that the distribution of these critical points is the most relevant to the underlying hypothesis.
	Notice particularly that the packing fraction of the largest discontinuities appears to be approaching the packing fraction of the lower end of the phase-coexistence interval with increasing $n$.
	Unfortunately, even using all of the techniques in Sec.\ \ref{subsec:graph_network}, the memory requirements increase so rapidly with $n$ that we could not realistically examine the spaces for $n \geq 7$.
	Even for five and six hard spheres, the volume of the space increases so rapidly with decreasing packing fraction that we cannot report diffusion diameters and $\epsilon$-mixing times over the entire domain.
	This does not substantially affect our conclusions though, since the sampled domain of $\eta$ already extends well below the coexistence interval, and notably includes all of the index-$1$ critical points.
	
\begin{figure}
		\centering
		\includegraphics[width=1.0\columnwidth]{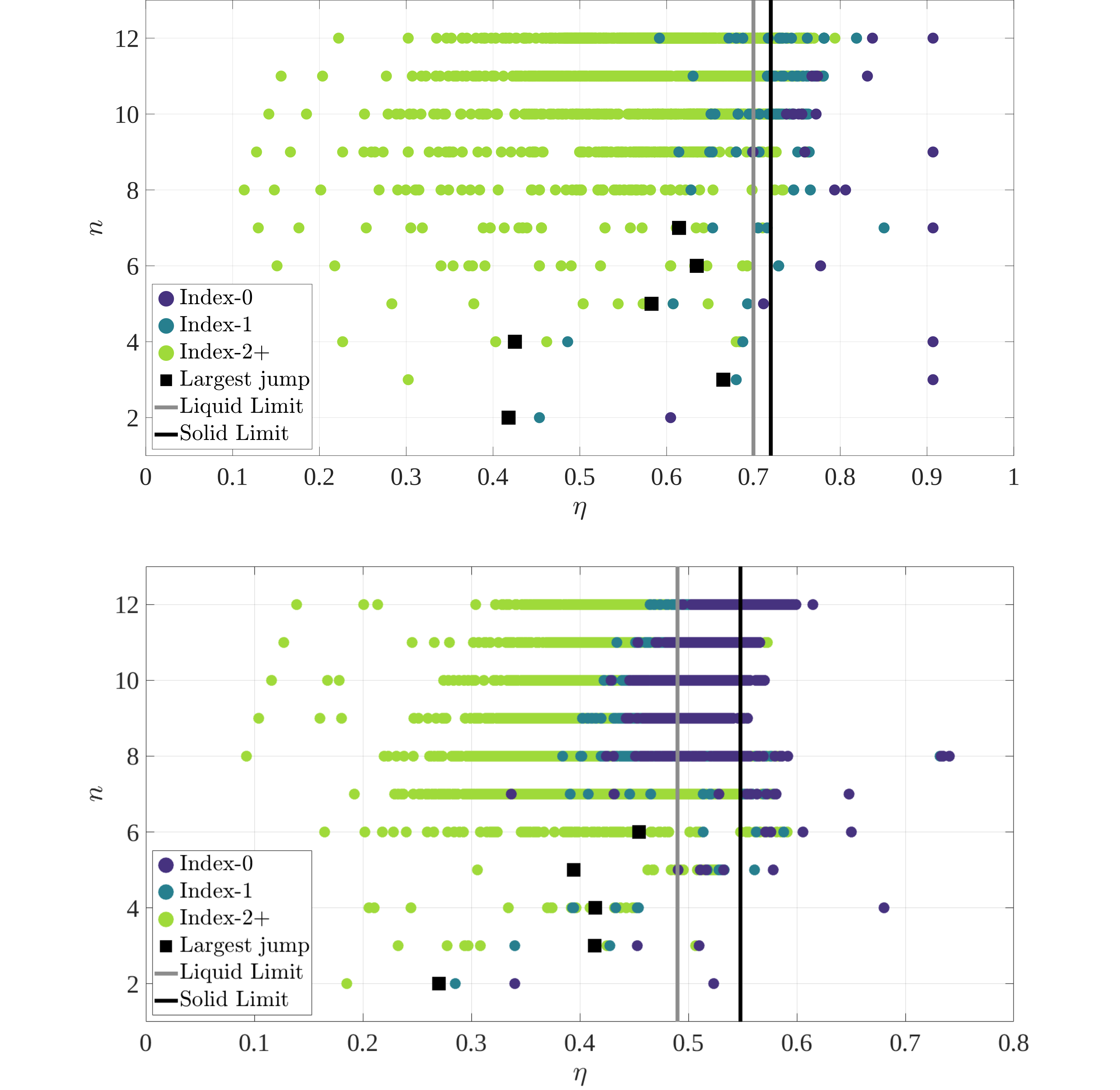}
		\caption{The largest discontinuous jumps (black squares) observed in Figs.\ \ref{fig:figure7} (top) and \ref{fig:figure8} (bottom) along with the packing fractions of all known critical points, colored by their indices (index-$0$ in purple, index-$1$ in blue, all others in green).}
		\label{fig:figure9}
	\end{figure}

	There is evidence that, at least for the hard disk systems, the substantial geometric changes leading to discontinuities in the $\epsilon$-mixing time are often associated with the lowest-packing-fraction index-$1$ critical point.
	Figure \ref{fig:figure9} shows the packing fractions of the largest discontinuities in Figs.\ \ref{fig:figure7} and \ref{fig:figure8} as black squares, with the number of hard disks (top) and hard spheres (bottom) increasing on the vertical axes.
	The packing fractions of all known critical points are also shown, with index-$0$ critical points in purple, index-$1$ critical points in blue, and all others in green.
	While $\epsilon$-mixing time data is only available for $n \leq 7$ hard disks and $n \leq 6$ hard spheres, populations of critical points for $n \leq 12$ hard disks and hard spheres have been sampled using established techniques that are described elsewhere \cite{ritcheyphd,ericok2021quotient,ericok2021configuration}.
	Apart from the increasing concentration of low-index critical points around the phase coexistence interval with increasing $n$, the most striking aspect of the figure is that the largest discontinuities almost always occur close to the packing fraction of the last index-$1$ critical point to appear with decreasing packing fraction (each of the black squares comes slightly before an index-$1$ critical point due to finite sampling).
	This is not entirely unexpected, since each index-$1$ critical point either joins previously disconnected components or add a new class of closed paths to the space.
	Supposing that discontinuities in the $\epsilon$-mixing time are associated with the former, it also makes sense that the largest discontinuity would be observed at lower packing fractions where the disconnected components being joined had the opportunity to grow to substantial volumes.
	The increasing concentration of low-index critical points around the phase coexistence interval draws the discontinuity closer to the liquid limit with increasing $n$, though that is admittedly a noisy trend for the small numbers of disks considered here.
	Nevertheless, this does provide evidence that something along the lines of Conj.\ \ref{conj:phase_transition} could be true for hard sphere systems, and perhaps for thermodynamics systems as well.
	
	\section{Conclusion}
	\label{sec:conclusion}
	
	A phase transition is necessarily related to a discontinuous change in the probability density function describing the distribution of the system's microstates on the phase space.
	The question of how this could occur as the thermodynamic control variables are continuously varied has not been conclusively answered in the literature.
	One proposal called the topological hypothesis \cite{caiani1997geometry,franzosi2004theorem} suggests that topological changes to the accessible region of the configuration space is a necessary condition for a phase transition.
	This paper instead suggests that a substantial change to the geometry of the accessible region is a necessary condition for a phase transition, and that such a change is often (but not always) associated with a topological change.
	More specifically, Conj.\ \ref{conj:phase_transition} proposes that a discontinuity in the mixing time on the configuration space is a necessary condition for a first-order phase transition in the thermodynamic limit.
	Our main result is a preliminary test of this conjecture for hard disk and hard sphere systems with few enough disks that the configuration space geometry can be explicitly studied.
	
	The configuration spaces of hard disks and hard spheres are represented as graphs with vertices representing specific configurations of disks and edges indicating the distances between the configurations. 
	The vertices of the graph are sampled more densely around critical points, or locations where the topology of the configuration space is known to change as a function of disk radius, to more accurately capture the topological and geometric changes in those regions.
	The diffusion diameters and the $\epsilon$-mixing times of the resulting graphs are calculated as functions of packing fraction for $n \leq 7$ and $n \leq 6$ hard spheres;
	the $\epsilon$-mixing time is proposed here, and is designed to be an analogue to the thermodynamic mixing time that can be explicitly evaluated.
	
	The geometric signals obtained for the diffusion diameter and the $\epsilon$-mixing time are consistent in the sense that their discontinuities generally occur at the same packing fractions.
	Apart from relating more directly to the content of Conj.\ \ref{conj:phase_transition}, the $\epsilon$-mixing time is a much smoother function of packing fraction than the diffusion diameter.
	The discontinuities in the $\epsilon$-mixing time are found to occur at comparatively few critical points; for the hard disk and hard sphere systems at least these are predominantly index-$1$ critical points.
	Along with the observation that the low-index critical points are increasingly concentrated around the phase coexistence interval with the number of disks, this suggests that a discontinuous change in the $\epsilon$-mixing time could indeed coincide with the first-order phase transitions in the hard disk and hard sphere systems in the thermodynamic limit.
	
	Future studies along these lines would likely need to extend the analysis to larger numbers of disks to make the trends in the approach to the thermodynamic limit clearer.
	Given the rate of increase in the computational requirements with number of disks, this would require at a minimum more efficient algorithms to search for critical points and calculate the distances between configurations.
	The computational requirements would also be reduced by using a further-reduced representation of the spaces by means of witness graphs \cite{aronov2013894}, or by restricting to particular intervals of disk radius close to the phase coexistence region.
	Further development of a suitable min-type Morse theory would also be helpful \cite{mori2011discrete,baryshnikov2014min}.

	\begin{acknowledgments}
		O.B.E.\ and J.K.M.\ were supported  by  the  National  Science  Foundation under Grant No.\ 1839370.
	\end{acknowledgments}

	\appendix

	\section{Heuristic for the distance}
	\label{sec:heuristic}
	
	Let $\mathcal{S} = \mathcal{T} \cup \mathcal{P} \cup \mathcal{L}$ in Eq.\ \ref{eq:distance_quotient}.
	Since $\mathcal{T}$ is a continuous group and $\mathcal{P} \cup \mathcal{L}$ is discrete, Eq.\ \ref{eq:distance_quotient} can be rewritten as
	\begin{align*}
		d_{\Lambda/\mathcal{S}}(\config{x},\config{y}) &= \min_{S \in \mathcal{P} \cup \mathcal{L} } d_{\Lambda/\mathcal{T}}[\config{x},S(\config{y})] \\
		d_{\Lambda/\mathcal{T}}[\config{x},S(\config{y})] &= \inf_{\vec{t} \in \mathcal{T}} d_{\Lambda}[\config{x}, S(\config{y}) + \config{t}] \\
		d_{\Lambda}[\config{x},S(\config{y})+\config{t}] &= \sum_{i = 1}^{n} \lVert \vec{x}_i - S(\vec{y}_i) - \vec{t} \rVert.
		\label{eq:distance_quotient_decomposed}
	\end{align*}
	Evaluating $d_{\Lambda/\mathcal{T}}[\config{x},S(\config{y})]$ involves solving a global optimization problem over rigid translations for a fixed discrete symmetry operation $S$.
	Since evaluating $d_{\Lambda/\mathcal{S}}(\config{x},\config{y})$ involves solving $n! \times O(\mathcal{L})$ of these global optimization problems, the computational cost could be significantly reduced if a suitable approximation for $d_{\Lambda/\mathcal{T}}[\config{x},S(\config{y})]$ could be found to quickly reject some of the $S$.
	
	Such an approximation is constructed by, for a fixed number of iterations $m$, randomly sampling a random translation $\config{t}$, calculating the distance $d_{\Lambda}[\config{x},S(\config{y})+\config{t}]$ for that translation, and accepting that translation only if it reduces $d_{\Lambda}[\config{x},S(\config{y})+\config{t}]$ \cite{wells2004generalized}.
	The resulting approximations for $d_{\Lambda/\mathcal{T}}[\config{x}, S(\config{y})]$ are sorted by increasing magnitude, and the full optimization problem is solved only for the first $M$ symmetry operations in the sorted list.
	The rationale for this procedure is that calculating the approximations using a moderate $m$ is less expensive than solving the global optimization problem.
	
	\begin{figure}
		\centering
		\includegraphics[width=1.0\columnwidth]{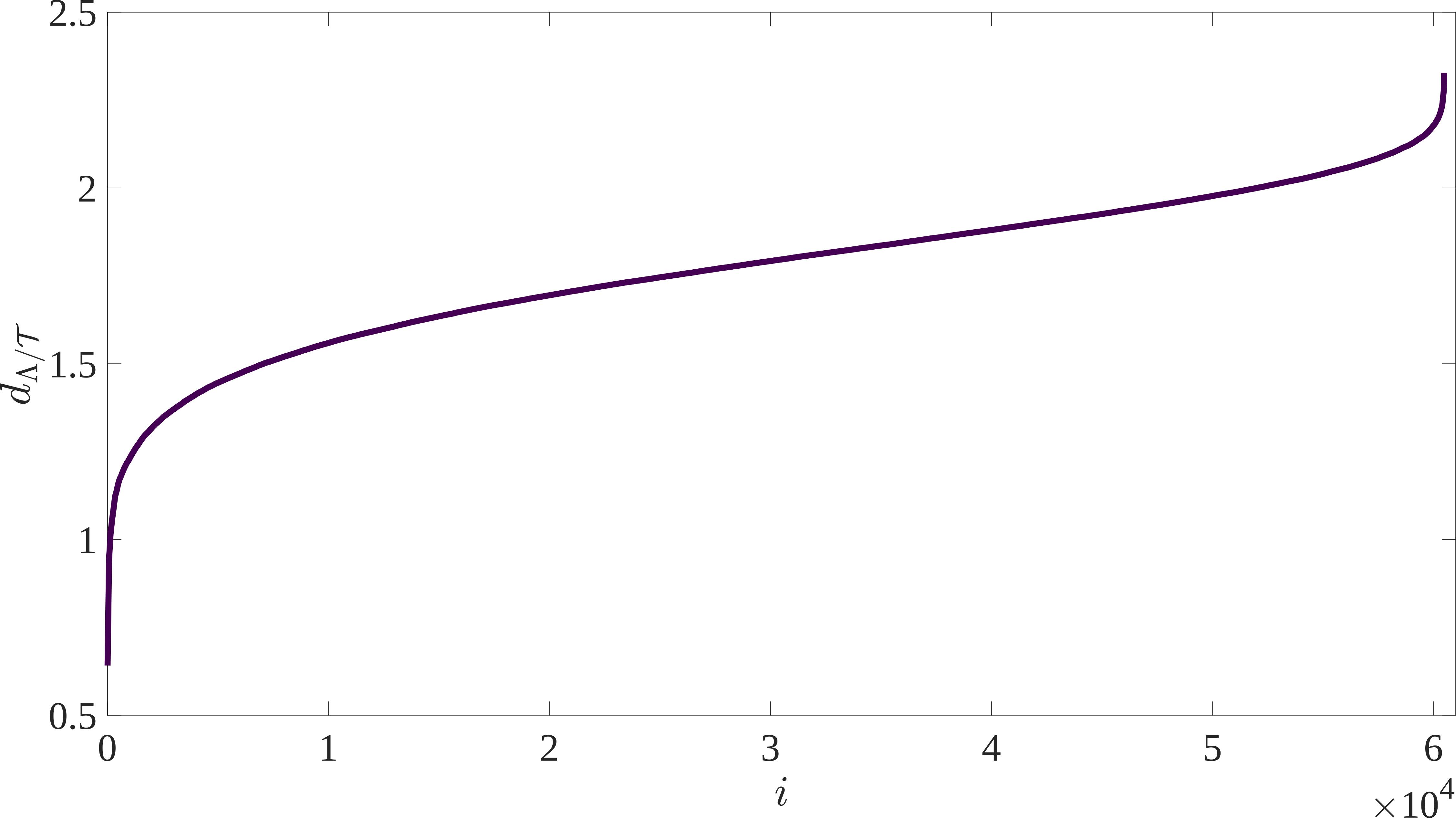}
		\caption{The sorted true distances $d_{\Lambda/\mathcal{T}}$ calculated for a fixed generic configuration and the $7! \times 12$ symmetric versions of a second generic configuration for $7$ hard disks.}
		\label{fig:figure10}
	\end{figure}
	
\begin{figure*}
		\centering
		\includegraphics[width=1.0\textwidth]{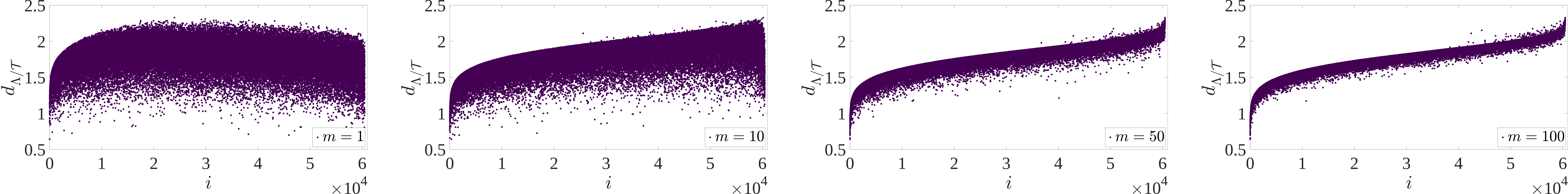}
		\caption{The same configurations as in Fig.\ \ref{fig:figure10} are sorted by the approximate values of $d_{\Lambda/\mathcal{T}}[\config{x},S(\config{y})]$ for $m=1$, $10$, $50$, and $100$ and are plotted with the true values of $d_{\Lambda/\mathcal{T}}[\config{x},S(\config{y})]$ on the vertical axis.} 
		\label{fig:figure11}
	\end{figure*}

	For two generic configurations of $7$ hard disks there are $7! \times 12 = 60\, 480$ symmetric versions of the second configuration due to the actions of $\mathcal{P}$ and $\mathcal{L}$.
	Figure \ref{fig:figure10} shows the sorted true distances $d_{\Lambda/\mathcal{T}}$, and $d_{\Lambda/\mathcal{S}}$ is the minimum of this set.
	Figure \ref{fig:figure11} shows the same configurations sorted by the approximate values of $d_{\Lambda/\mathcal{T}}[\config{x},S(\config{y})]$ for $m=1$, $10$, $50$, and $100$ plotted with the true values of $d_{\Lambda/\mathcal{T}}[\config{x},S(\config{y})]$ on the vertical axis.
	Observe that with increasing $m$ these curves should converge to the one in Fig.\ \ref{fig:figure10}, and that even for small $m$ the true minimum distance should appear within the first $M \ll 60\, 480$ symmetry operations.
	Numerical experiments suggest that $m = 50$ and $M = 200$ are generally sufficient for an approximation with a relative error on the order of $10^{-4}$.
	
	\section{Eigenspectrum analysis}
	\label{sec:eigenspectrum_analysis}
	
\begin{figure}
		\centering
		\includegraphics[width=1.0\columnwidth]{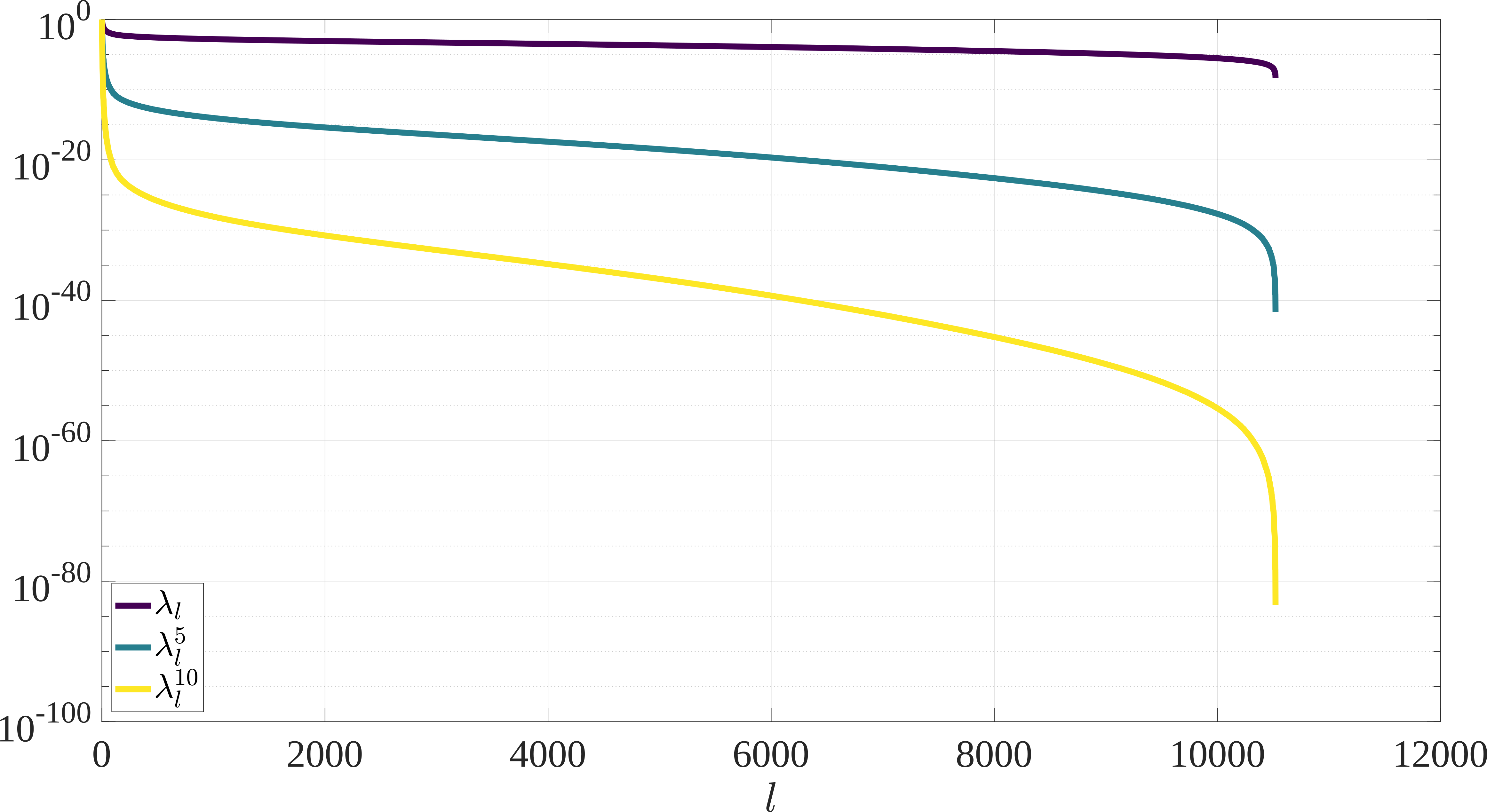}
		\caption{The $t$th powers of the eigenvalues of $\mat{P}$ for $\Lambda(4)/\{\mathcal{T} \cup \mathcal{P} \cup \mathcal{L}\}$ at times $t = 1$, $5$, and $10$.
			The values generally decay rapidly, and are often numerically indistinguishable from zero for sufficiently large $t$.}
		\label{fig:figure12}
	\end{figure}

	The diffusion coordinates in Eq.\ \ref{eq:diffusion_maps} involve the $t$th powers of the eigenvalues of $\mat{P}$ sorted by decreasing magnitude.
	This means that for sufficiently large $t$, the computational expense of calculating the diffusion distance can be considerably reduced with negligible loss of accuracy by truncating the eigenspectrum and only considering the first few diffusion coordinates \cite{talmon2013diffusion}.
	Figure \ref{fig:figure12} shows the $t$th powers of the eigenvalues of $\mat{P}$ for $\Lambda(4)/\{\mathcal{T} \cup \mathcal{P} \cup \mathcal{L}\}$ at times $t = 1$, $5$, and $10$.
	Observe that even for the relatively small $t = 5$ most of the diffusion coordinates will be numerically indistinguishable from zero.
	Only the first $200$ diffusion coordinates are used here unless otherwise specified, changing the diffusion diameter to a relative precision of less than $10^{-8}$ and making this source of error negligible compared to other sources.
	
	\section{Mixing time of an example system}
	\label{sec:mixing_time_toy}
	
\begin{figure}
		\centering
		\includegraphics[width=1.0\columnwidth]{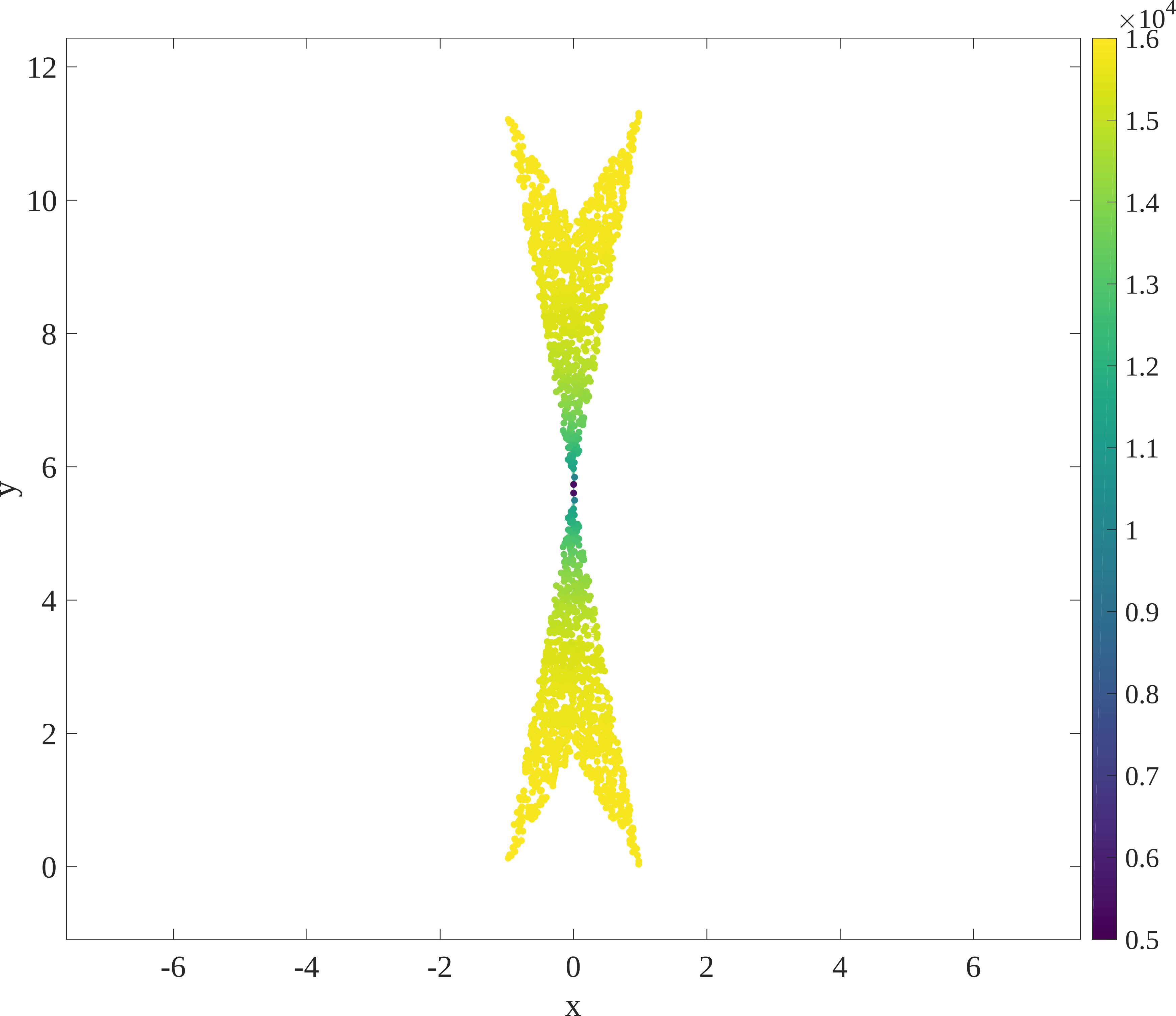}
		\caption{Individual mixing times $t_\epsilon(\vec{q}_0)$ of the example system shown in Fig.\ \ref{fig:figure1}.
			The average mixing time $\langle t \rangle_\epsilon$ is around $15\ 000$.}
		\label{fig:figure13}
	\end{figure}

	Figure \ref{fig:figure13} shows the $\epsilon$-mixing times $t_\epsilon(\vec{q}_0)$ as a function of $\vec{q}_0$ for the example system in Fig.\ \ref{fig:figure1}.
	The $k$-nearest neighbor graph is constructed with $k = 7$, the length of an edge is given by the Euclidean distance between the corresponding vertices, and $\epsilon = 0.001$ is used since the steady-state probability mass at each vertex is $0.0005$ (the inverse of the number of vertices).
	Points closer to the center have smaller $\epsilon$-mixing times since the initial Dirac-delta distribution can spread to the upper and lower regions more easily as compared to distributions starting near the corners.
	The average mixing time $\langle t \rangle_\epsilon$ for this example is around $15\ 000$.
	
\end{document}